\documentstyle[12pt,moriond]{article}
\def\spose#1{\hbox to 0pt{#1\hss}}
\def\lta{\mathrel{\spose{\lower 3pt\hbox{$\mathchar"218$}}
     \raise 2.0pt\hbox{$\mathchar"13C$}}}
\def\gta{\mathrel{\spose{\lower 3pt\hbox{$\mathchar"218$}}
     \raise 2.0pt\hbox{$\mathchar"13E$}}}
\def\etal{{\it et al.\ }}
\def\ni{\noindent}

\begin{document}
\heading{DWARF GALAXIES OF TIDAL ORIGIN -- RELEVANT FOR COSMOLOGY ?}

\author{U. Fritze -- v. Alvensleben $^{1}$, C. S. M\"oller $^{1}$, P. - A. Duc $^{2}$
} {$^{1}$ Universit\"atssternwarte G\"ottingen, Germany.}  {$^{2}$ ESO Garching, Germany.}

\begin{moriondabstract}
Evolutionary synthesis models for Tidal Dwarf Galaxies ({\bf TDG}s) are presented
that allow to have varying proportions of young stars formed in the
merger-induced starburst and of stars from the merging spirals' disks. 
The specific metallicities as well as the gaseous emission of actively star 
forming TDGs are consistently accounted for. Comparison of 
models with observational data (e.g. Duc, {\sl this volume}) 
gives information on the present evolutionary state and possible 
future luminosity evolution of TDGs. The redshift 
evolution of merger rates and of the gas content and metallicities of spiral galaxies are used to 
estimate the number of TDGs at various redshifts and to investigate their 
contribution to magnitude limited surveys. 

\end{moriondabstract}

\section{Properties of nearby TDGs}
\subsection{Metallicities}
As shown by P.-A. Duc ({\it this volume}), metallicities of nearby TDGs are higher than predicted by the 
L -- Z - relation for dwarfs (Skillman \etal 1989). Rather they are in the range of observed spiral galaxy 
ISM abundances. The metallicity is important since it both determines 
the peak brightness as well as the fading after a burst of given maximum star formation rate ({\bf SFR}). 
E.g., for ${\rm Z_{\odot} \longrightarrow Z=0.001}$ the peak brightness of a given burst increases by 
$\sim 1$ mag in B, V, and K.  

\subsection{Stellar Populations}
Presently, numerical simulations yield two formation scenarii for TDGs. 
While Barnes \& Hernquist (1992) find stellar 
condensations along stellar 
tidal tails comprising the composite stellar population of the parent spiral disk with a large mass scale 
$1 \cdot 10^7 - 4 \cdot 10^8 {\rm M_{\odot}}$, Elmegreen \etal (1993) report gaseous condensations along 
gaseous tails with a mass scale set by the encounter (up to a factor $\sim 5$) which may lead to young 
starburst populations for TDGs. 

\ni
In the framework of our evolutionary synthesis modelling we describe TDGs with two ingredients in varying 
proportions to cover both possible formation scenarii: \\
\hspace{.5truecm} -- a composite stellar population torn out from the disk of an interacting spiral \hfill\break
\hspace{.5truecm} -- a starburst taking place in the tidal debris. 
 
\ni
We assume the same Scalo IMF from 0.1 -- 85 M$_{\odot}$ both for the progenitor spiral and the starburst for simplicity. 
We use recent stellar evolutionary tracks from the Geneva group, Lyman continuum photons from 
Schaerer \& de Koter (1997) and emission line ratios relative to ${\rm H_{\beta}}$ from Stasi\'nska (1984) for 
${\rm Z_{\odot}}$ and Izotov \etal (1994)  for ${\rm Z < Z_{\odot}}$ to include both the gaseous continuum and 
line emission. Star formation histories for spiral galaxies of various spectral types are adopted from our 
spectrophotometric modelling of undisturbed galaxies (Fritze -- v. Alvensleben \& Gerhard 1994a, b),  starbursts are 
described by exponentially declining SFRs with e-folding times ${\rm t_{\ast}}$ 
in the range $1 \cdot 10^6 - 1 \cdot 10^8$ yr. 
The age of the spiral before the interaction is assumed to be $\sim 12$ Gyr for nearby TDGs, the 
relative fractions of ''old'' disk and young burst stars determine the burst strengths. 

\ni
Comparing observed properties of P.-A. Duc's TDG sample to a small grid of models with various burst strengths, 
metallicities ${\rm Z \geq Z_{ISM}^{spiral}}$, and burst timescales ${\rm t_{\ast}}$ yields the following results:
\begin{itemize}
\item Most TDGs seem to feature quite strong starbursts (${\rm  S^{young} / S^{old} = 0.10 - 0.40}$ with S being 
the stellar mass of the young and old populations, respectively.
\item Burst SF timescales are ${\rm 10^6~yr \lta t_{\ast} \lta 10^7~yr}$.
\item For typical metallicities of nearby TDGs, fading after a strong burst amounts to 
3.3 mag, 2.8 mag, 2.4 mag, and 1.2 mag in B, V, R, and K, 
respectively, within the first Gyr. For lower metallicities, fading is slightly weaker. 
\item Gaseous emission is important for broad band colours in starbursting TDGs.
\end{itemize}
\ni
Due to uncertainties in the internal reddening corrections (${\rm E_{B-V}}$ from the Balmer decrement might overestimate 
internal reddening of broad band colours) it is not clear at present if there are ``purely old'' or pure burst objects 
within the nearby TDG sample. Those objects would have much weaker (purely old) or stronger fading (pure bursts) 
than the strong burst models.  

Detailed analysis has only been performed for three TDGs. Object A105S in the stellar and HI rich tidal tail of Arp 105 
Duc \& Mirabel 1994) has 
a strong burst of age $(2 - 6) \cdot 10^7$ yr and a short ${\rm t_{\ast} \sim 10^6}$ yr. The ``old'' population from the 
spiral disk contributes 4 \% to the light in B, 44 \% to the light in K, and $\sim 75$ \% to the total 
stellar mass of this TDG. Objects N5291a and N5291i around the early type galaxy NGC 5291 with merger signature and 
${\rm \sim 5 \cdot 10^{10} M_{\odot}}$ of HI accreted from a gas-rich galaxy (Duc \& Mirabel 1998) but without any detectable 
{\bf stellar} tidal tail are compatible with ``pure burst'' models of young age $<10^6$ yr and short ${\rm t_{\ast} \sim 
10^6}$ yr as well as with a strong burst of age ${\rm < few 10^7}$ yr, ${\rm t_{\ast} \lta 10^7}$ yr that might hide 
an ``old'' population containing as much as 10 times the mass of the current starburst. Fading of N5291a and N5291i 
crucially depends on the presence or absence of this ``old'' population. 

\section{Clues to the Cosmological Significance of TDGs} 
Merger rates increasing drastically with redshift, an obvious question is in how far TDGs might contribute to the faint 
galaxy counts of dwarf galaxies. Several attempts have been made to explain 
the faint end of the Luminosity Function by a fading population of dwarf galaxies. Here, we lean on the scenario 
of Babul \& Ferguson (1996) who envisage formation of a population of dwarf galaxies delayed by a strong  
ionising UV background until ${\rm z \sim 1}$. They find that a population of dwarf galaxies of ${\rm 10^9~M_{\odot}}$, 
${\rm Z \sim \frac{1}{20} \cdot Z_{\odot}}$ with a burst SFR of 3 ${\rm M_{\odot}yr^{-1}}$ and ${\rm t_{\ast} \sim 10^7}$ yr, if 
formed over a timespan of ${\sim 2 \cdot 10^9}$ yr after ${\rm z = 1}$, may dominate the counts at 
${\rm B > 25~ mag}$ without significantly contributing to the I and K counts and also agree with the redshift distribution 
of the Faint Blue Galaxies ({\bf FBG}s). Their problem was with the remnants of this fading dwarf population, since -- 
locally -- dSphs do show stellar components formed earlier than ${\rm z = 1}$ and LSB remnants tend to be gas-rich and not 
gas-free as in their scenario. 

The aim of the present investigation is to show that  -- if numerous enough -- TDGs might do the job of 
Babul \& Ferguson's hypothetic dwarf galaxy population with the advantage of largely avoiding the remnant problem: 
If either they use up or blow away their gas, the stellar content of TDGs containing an ``old'' disk population 
plus a starburst closely resembles that of some nearby dSphs (e.g. Grebel 1997), allthough a large DM content of dSphs -- 
highly controversial at present -- would be a serious problem. 
Many TDGs have ample HI and if they keep 
their gas while their burst fades they may come to resemble LSBs. Moreover, fall back onto the merger remnant 
may keep part of the intermediate to high redshift TDG population from still being around locally.  

The interaction rate of field galaxies is known to steeply increase with redshift. The uncertainty of Zepf \& Koo's 
(1989) estimate ${\rm \sim (1+z)^{4 \pm 2.5}}$ probably still encompasses the range discussed in more recent determinations, 
also from the HDF. Our spiral galaxy models show that ISM metallicities decrease with redshift as 
${\rm Z \sim (1+z)^{\zeta}}$ with $\zeta = 
-2.0$ for late type spirals Sc, Sd and  $\zeta = -1.6$ for Sb spirals in the redshift range ${\rm z \lta 2}$, making 
starbursts brighter in the past. At the same time, the gas content of spirals increases significantly 
with redshift already at redshifts ${\rm z \lta 0.5}$. Our models indicate 
${\rm G/M \sim (1+z)^{\gamma}}$ with $\gamma = 1.2 ~(3.2)$ for spirals of type Sc, Sd (Sb) (Fritze -- v. Alvensleben 1995). 
Moreover, gaseous disks 
probably were more extended in the past, allowing for stronger gaseous tidal features and numerous TDGs. A possible 
correlation between the gas content and the number of TDGs typically formed could be derived from a larger sample of 
nearby interacting galaxies. The observed metallicity range of nearby TDGs corresponds to the range of typical ISM 
abundances from Sbc through Sdm galaxies (cf. Zaritsky \etal 1994), their upper metallicity limit might indicate 
that Sa galaxies do not have enough gas to form bright \& starbursting TDGs. A hypothetical TDG forming in an 
Sd (Sc) tidal arm at ${\rm z \sim 0.5~(z \sim 1)}$ is predicted by our models to have ${\rm 12 + log(O/H) = 7.9}$, a value 
typical for a dIrr of ${\rm M_B \sim -15}$ mag, i.e., it would no longer be conspicuous in metallicity. 

Typical luminosities of nearby TDGs lie in the range $- 15 \dots - 16$ mag. Together with bolometric distance moduli 
for ${\rm H_0 = 50,~\Omega_0 =1}$ and chemically consistent evolutionary and cosmological corrections ${\rm(e + k)_B}$ 
as derived for starburst galaxies of the appropriate metallicities, magnitudes ${\rm B \sim 26}$, 27, 28 
are predicted for TDGs at ${\rm z = 0.3,~ 0.6,~ 1.0}$, respectively. I.e., over the range ${\rm 24 \lta B \lta 29}$ mag where 
the excess population of FBGs is observed, TDGs might contribute from a redshift range ${\rm z \lta 0.1}$ to 
${\rm z \gta 1.0}$. 

Until ${\rm z = 0}$, a typical TDG at redshift ${\rm z = 0.05 ~(0.3)}$ will have faded by $2.5 \pm .5~ (4.3 \pm .8)$ 
mag in B and by $\sim 1$ mag in K for the strong burst case, and by as much as $4 \pm 1~(6 \pm 1)$ mag in B and 
$2.5 \pm .5 ~(4 \pm 1)$ mag in K for the ``pure burst'' case. Errors reflect uncertainties in ${\rm t_{\ast}}$ 
and [O/H]. This fading in B is sufficient for the Babul \& Ferguson model, the fading in K crucially 
depends on the presence or absence of an ``old'' population of stars from an interacting galaxy. 

As shown by Hibbard \& Mihos (1996) in their dynamical modelling accounting for the precise HI velocity structure along 
the tidal tails of the merger remnant NGC 7252, about ${\rm 50~ \%}$ of the material along those tails is bound to fall back 
onto the merger remnant on timescales of a few Gyr. If the velocity structure in the NGC 7252 tails is somehow typical 
for interacting systems this fall back will also occur to any TDGs formed within the lower part of tidal tails. Moreover, 
TDGs being formed from disk material, they won't have dark matter halos to stabilize against tidal disruption by the parent 
galaxy or against SN driven winds from their own strong starbursts. If these effects can disrupt the TDG or not will 
depend on whether or not an ``old'' underlying population of stars contributes to the potential depth.  
So, in summary, we expect less -- and probably much less -- than 50 \% of the original TDG population to 
survive, and those tend to be the brightest ones often seen at the tips of tidal tails. These may well contain an 
important, perhaps even dominant, ``old'' stellar population and may or may not have retained some part of the 
ample HI supply generally observed in local TDGs. Thus, the remnants of cosmological TDGs might look like local dEs, dSphs, dIrrs, 
BCDs or even the LMC. We recall that -- contrary to nearby examples -- TDGs from ${\rm z \gta 0.3}$ will not stand 
out in [O/H] above the normal dwarf galaxy population. 

\section{Conclusions and Outlook}
The number of TDGs is expected to strongly increase with redshift. They will strongly fade in B and many of them 
will disappear until ${\rm z \sim 0}$. Thus, TDGs might well account for some part of the FBG excess. To work out 
the details, we need more data and modelling of nearby TDGs. In particular, we need to know if they typically 
contain an ``old'' population of stars and how much of it, and in how far the number of TDGs per interaction and their 
properties depend on the parameters of the encounter, as e.g. the gas content of the interacting galaxy. It would also be 
very useful to have some dynamical modelling of TDGs within the tidal field of their parent galaxies.


\begin{moriondbib}
\bibitem{} Babul, A., Ferguson, H.~C., 1996, \apj {458} {100}
\bibitem{} Barnes, J.~E., Hernquist, L., 1992, \nat {360} {715}
\bibitem{} Duc, P.-A., Mirabel, I.~F., 1994, \aa {289} {83}
\bibitem{} Duc, P.-A., Mirabel, I.~F., 1998, \aa {333} {813}
\bibitem{} Elmegreen, B.~G., Kaufman, M., Thomasson, M., 1993, \apj 
        {412} {90}
\bibitem{} Fritze -- v. Alvensleben, U., in {\it QSO Absorption Lines}, ed. G. Meylan, Springer, p. 81
\bibitem{} Fritze -- v. Alvensleben, U., Gerhard, O.~E., 1994a, b \aa 
   	{285} {751} $+$ {775}
\bibitem{} Grebel, E.~K., 1997, {\it Rev. Mod. Astron.} {\bf 10}, 29
\bibitem{} Hibbard, J.~E., Mihos, J.~C., 1996, \aj {111} {655}
\bibitem{} Izotov, Y.~I., Thuan, T.~X., Lipovetsky, V.~A., 1994, \apj 
        {435} {647}
\bibitem{} Schaerer, D., de Koter, A., 1997, \aa {322} {598}
\bibitem{} Skillman, E.~D., Kennicutt, R.~C., Hodge, P.~W., 1989, \apj {347} {875}
\bibitem{} Stasi\'nska, G., 1984, \aas {55} {15}
\bibitem{} Zaritsky, D., Kennicutt, R.~C., Huchra, J.~P., 1994, \apj {420} {87}
\bibitem{} Zepf, S.~E., Koo, D.~C., 1989, \apj {337} {34}
\end{moriondbib}
\end{document}